\documentclass[onecolumn,numbers,sort&compress]{revtex4-2}
\usepackage[utf8]{inputenc}
\usepackage{amsmath, amssymb}
\usepackage{aas_macros}
\usepackage{graphicx}
\usepackage{subcaption}
\usepackage{caption}
\usepackage{url}
\usepackage{hyperref}
\hypersetup{
    colorlinks=true,
    linkcolor=blue,
    filecolor=magenta,      
    urlcolor=cyan,
    citecolor=green
    }
\usepackage{float}
\usepackage{booktabs}
\usepackage{fontawesome5}
\usepackage{multirow}
\usepackage{comment}
\usepackage{tikz}
\usetikzlibrary{positioning,arrows.meta,fit}

\begin{document}

\title{Bayesian Inference of Primordial Magnetic Field Parameters from CMB with Spherical Graph Neural Networks\\}
\author{Juan Alejandro Pinto Castro$^{1}$}
\author{ H\'ector J. Hort\'ua$^{1,2}$ }
\author{Jorge Enrique Garc\'ia-Farieta$^{1}$}
\author{Roger Anderson Hurtado$^{3}$}
\affiliation{$^{1}$Laboratorio de Inteligencia Artificial (SavIA Lab), Grupo Signos, Departamento de Matemáticas, Universidad El Bosque, Bogotá,
Colombia}
\affiliation{$^{2}$Instituto de Neurociencias, Universidad El Bosque, Bogotá, Colombia}

\affiliation{$^{3}$Observatorio Astronómico Nacional, Universidad Nacional de Colombia, Bogotá, Colombia}

\begin{abstract}

Deep learning has emerged as a transformative methodology in modern cosmology, providing powerful tools to extract meaningful physical information from complex astronomical datasets. This paper implements a novel Bayesian graph deep learning framework for estimating key cosmological parameters in a primordial magnetic field (PMF) cosmology directly from simulated Cosmic Microwave Background (CMB) maps. Our methodology utilizes \texttt{DeepSphere}, a spherical convolutional neural network architecture specifically designed to respect the spherical geometry of CMB data through \texttt{HEALPix}  pixelization. To advance beyond deterministic point estimates and enable robust uncertainty quantification, we integrate Bayesian Neural Networks (\texttt{BNNs}) into the framework, capturing aleatoric and epistemic uncertainties that reflect the model confidence in its predictions.
The proposed approach demonstrates exceptional performance, achieving  $R^{2}$ scores exceeding 0.89 for the magnetic parameter estimation. We further obtain well-calibrated uncertainty estimates through post-hoc training techniques including Variance Scaling and GPNormal. This integrated DeepSphere-\texttt{BNNs} framework not only delivers accurate parameter estimation from CMB maps with PMF contributions but also provides reliable uncertainty quantification, providing the necessary tools for robust cosmological inference in the era of precision cosmology~\href{https://github.com/JavierOrjuela/Bayesian-SphericalCNN}{\faGithub}.

\end{abstract} 

\maketitle

\section{Introduction}
Cosmic Microwave Background (CMB) radiation is a cornerstone of modern cosmology as it provides a snapshot of the physics of the early Universe. The CMB is sensitive to physical densities because the evolution of acoustic oscillations in the primordial plasma depends on them directly, not just on their fractional contribution to the total energy density \cite{dodelson2003}. Precise estimation of key cosmological parameters such as the physical baryon density $\omega_b$, cold dark matter density $\omega_c$, reionization optical depth $\tau$, scalar spectral index $n_s$, the amplitude of the primordial power spectrum $A_s$, and Hubble parameter $H_0$, is essential for understanding the composition and evolution of the Universe. In particular, the Hubble parameter, which describes the expansion rate of the Universe, has garnered significant attention due to the so-called Hubble tension. This term refers to a persistent and statistically significant discrepancy between the values derived from independent cosmological measurements of the early Universe and those obtained from local data. 
Primordial Magnetic Fields (PMFs) offer a potential solution to the Hubble tension that is not inherently fine-tuned for that purpose. Previous studies such as \cite{2021PhRvD.104j3517R, 2025JCAP...03..012J, 2025arXiv250616517S} have shown that PMFs can induce baryon inhomogeneities, which accelerate the recombination process and consequently increase the CMB-inferred value of $H_0$, bringing it into closer agreement with local measurements. The PMF scenario is particularly compelling because it is grounded in well-established theoretical frameworks that predate the Hubble tension, having long been investigated as a possible origin of the large-scale magnetic fields observed in galaxies and clusters; see e.g. \cite{2002RvMP...74..775W, 2010Sci...328...73N, 2012SSRv..166...37W}.\\
Beyond their potential role in addressing cosmological tensions, PMFs serve as valuable probes of early-universe physics. They are motivated by high-energy processes, such as inflationary magnetogenesis and cosmological phase transitions. As such, constraints on PMFs derived from CMB observations offer a promising avenue for investigating the origin and evolution of cosmic magnetism, while also providing natural predictions in certain extensions of the Standard Model \cite{2013CaJPh..91..451P}. Furthermore, PMFs imprint distinct signatures on the CMB by sourcing scalar, vector, and tensor perturbations, thereby modifying both temperature and polarization anisotropies~\cite{Hort_a_2017}. PMFs can generate vector modes and B-mode polarization patterns that are absent in the standard $\Lambda$CDM model, as well as induce Faraday rotation of the CMB's polarization plane \cite{2011PhRvD..84d3530P}, producing a frequency-dependent signal that upcoming experiments, such as the Simons Observatory, LiteBIRD, and CMB-S4, aim to detect \cite{2017ApJ...846..164S}.
\\
On the other hand, the choice of inference methodology and statistical framework is a critical determinant of the precision and reliability of any constraints on cosmological parameters or PMF models~\cite{zucca2017}. Robust bounds depend on the accurate modeling of the likelihood, thorough treatment of systematics, well‑motivated priors, and a careful handling of parameter degeneracies.  In addition, traditional techniques applied to the power spectrum are computationally expensive and may not fully exploit the non-Gaussian information present in the CMB maps. This creates a compelling opportunity for deep learning techniques to provide faster and potentially more accurate estimates by learning directly from  high-resolution maps~\cite{hortua2020accelerating}.
\\
A primary challenge in applying standard deep learning architectures to CMB analysis is the data's inherent spherical geometry. Conventional convolutional neural networks, designed for flat grids, introduce distortions and are ill-suited for processing data on the sphere. From that perspective, this work is intended to address this issue using the Spherical convolutional neural network \texttt{DeepSphere}~\cite{perraudin2019}. \texttt{DeepSphere} uses a pixelation scheme to represent spherical data, called \texttt{HEALPix}, which divides the sphere into pixels of equal area in a hierarchical way, allowing the neural network to efficiently perform convolutions and analysis of the celestial sphere~\cite{Krachmalnicoff_2019}.
\\
Beyond geometric challenges, robust inference requires not only accurate predictions but also reliable uncertainty quantification. Standard deep learning models provide deterministic point estimates, failing to capture the epistemic uncertainty associated with limited data or model limitations. To this end, we integrate Bayesian Neural Networks (\texttt{BNNs}) into our architecture. \texttt{BNNs} learn a probability distribution over the model weights, naturally quantifying predictive uncertainty.
\\
This work is organized as follows: Section \ref{sec:2} provides background on Graph Neural Networks, DeepSphere, and Bayesian Neural Networks. Section \ref{sec:3} details our methods, including the simulation pipeline, data validation, network architecture, and training procedure. Section \ref{sec:4} presents the results and analysis of the model's performance and uncertainty calibration. Finally, Sections \ref{sec:5} and \ref{section6} conclude the paper and discuss future research directions.

\section{Neural Networks\label{sec:2}}
Deep learning has become an essential tool in modern cosmology, transcending the limitations of traditional statistical methods to address the challenges posed by the massive, complex, and high-dimensional nature of CMB \cite{Dvorkin2022, hortua2020,GarciaFarieta2024}. This research leverages convolutional operations extended to graph neural networks to predict cosmological parameters, specifically, we use the \texttt{DeepSphere} framework \cite{perraudin2019}. \texttt{DeepSphere} offers an efficient implementation of spherical convolutions by representing the celestial sphere as a graph and performing spectral or spatial graph convolutions, thus preserving the global geometry of the CMB data. Furthermore, we incorporate Bayesian Neural Networks which learn the mean and variance to define a probability distribution over the weights. This approach explicitly quantifies epistemic uncertainty, reflecting the model's confidence in its predictions arising from limited data or inherent model limitations \cite{Gal2016,Blundell2015}.

\subsection{Graph Neural Networks and DeepSphere}

Graph neural networks (\texttt{GNNs}) are a class of supervised Deep learning model~\cite{Farsian2022} which can learn functions over graphs and are a leading approach for building predictive models on graph-structured data $\mathcal{G}(\mathcal{V}, \mathcal{E})$ \cite{Corso2024}, where nodes $\mathcal{V}$ are objects interconnected by edges $\mathcal{E}$ which are relationships \cite{Bessadok2023}. \texttt{GNNs} represent a formidable progression in deep learning designed to process data structures in non-Euclidean domains where Convolutional Neural Networks (CNNs) have limitations \cite{Zhang2019,Xu2023}. The success of \texttt{GNNs} in handling arbitrary relational data has led to the broader field of geometric graphs, which aims to generalize deep learning operations to diverse non-Euclidean manifolds \cite{Bronstein2021}. In some domains, the nodes in the graph are embedded in the 3D space, coordinates could be used as features of the nodes \cite{Corso2024}. In cosmology, objects like dark matter halos and galaxy clusters are represented as nodes connected by edges that trace cosmic filaments and spatial proximity \cite{Farsian2022}. When working with geometrical graphs, it is important to consider the symmetry of the task with respect to translations and rotations of the frame of reference \cite{Corso2024}, and generalize the data defined on a curved surface, such as CMB maps with PMFs and weak lensing, this is the challenge, a method that respect spherical symmetry \cite{Dvorkin2022}. This challenge drives the development of Spherical Convolutional Neural Networks (SCNNs) such as \href{https://github.com/deepsphere}{\texttt{DeepSphere}}, which is computationally efficient and well adapted to the spherical geometry of the CMB. This adaptation is essential because the CMB is naturally represented on the two-sphere $\mathbb{S}^2$ and is typically pixelized using the \texttt{HEALPix} scheme \cite{gorski2005}. The \texttt{HEALPix} pixels are treated as nodes of a graph, and local neighborhoods are exploited through spherical convolutions defined directly on the pixelization. Many of the standard CNN operations, such as convolutions and pooling, with filters restricted to being radial, \texttt{DeepSphere} can be made it invariant or equivariant to rotation. This ensures that the extracted features are independent of the map's orientation, a necessary constraint for robust inference. \texttt{DeepSphere} has been shown to be computationally more efficient than harmonic-based methods and yields superior performance in tasks such as cosmological model discrimination, particularly in scenarios involving high noise levels or partial sky coverage, confirming its utility for next-generation surveys \cite{perraudin2019}.

\subsection{Bayesian Neural Networks}
Standard deep learning models are characterized by their deterministic nature, learning single-point estimates for all model parameters. While it is effective for prediction, those models fail to quantify the epistemic uncertainty inherent in the data and prediction process, a critical requirement for robust scientific inference \cite{Jospin2022}. Bayesian Neural Networks offer a principled approach to quantifying uncertainty in deep learning models. They treat the model parameters as random variables governed by probability distributions, thereby capturing epistemic uncertainty \cite{Magris2023}. Inference in \texttt{BNNs} involves computing the posterior distribution, $p(\mathbf{w}|\mathcal{D})$ over the weights, $\texttt{w}$, given the training data $\mathcal{D}$. This is commonly achieved via scalable methods such as Variance Inference or Monte Carlo Dropout \cite{Blundell2015, Gal2016}.  Variance inference is the preferred and scalable method for training \texttt{BNNs}, which finds a distribution $q(\mathbf{w}|\mathbf{\theta})$ that minimizes the difference to the true posterior $p(\mathbf{w}|\mathcal{D})$ \cite{Magris2023}  as well as a prior distribution on the weights $p(\mathbf{w})$. This step is reached by minimizing the KullBack-Leibler (\texttt{KL}) divergence between the two distributions, and is equivalent to maximizing the evidence lower bound (\texttt{ELBO}) \cite{Gal2016}, that includes a term associated with the negative log-likelihood (\texttt{NLL}), as \cite{hortua2020}
\begin{equation}
    \rm{ELBO}\sim~KL-\text{NLL}= KL(q(\mathbf{w}|\theta)||p(\mathbf{w}))- \big(\frac{1}{2}\log|\Sigma| + \frac{1}{2} (\textbf{y}- \mu)^{T}\Sigma^{-1}(\textbf{y}-\mu)\big),
    \label{nll}
\end{equation}
where $\Sigma$ is the predicted covariance matrix, $\textbf{y}$ is the observed data and $\mu$ is the predicted mean vector from the model. The application of \texttt{BNNs} to cosmological parameters estimation has been investigated in recent works \cite{hortua2020,GarciaFarieta2024}. The combination of \texttt{GNNs} and \texttt{BNNs} results in a powerful framework, which not only achieves high predictive accuracy, but crucially reliable estimates of epistemic uncertainty.
\section{Methods\label{sec:3}}

To analyze the simulated CMB with PMF maps and extract cosmological information, we employed \texttt{DeepSphere}, a graph-based convolutional neural network architecture specifically designed for spherical data \cite{perraudin2019}. Specifically, we make use of the \texttt{HealpyPseudoConv} layers, which implement convolution operations on the sphere by aggregating information from neighboring pixels according to the hierarchical structure of \texttt{HEALPix}. In addition, we employ \texttt{HealpyPooling} layers, which perform a coarsening of the sphere by reducing the resolution defined by \texttt{nside}. This combination of spherical convolution and pooling layers allows the network to efficiently capture both large-scale and small-scale anisotropy patterns of the CMB, while remaining computationally feasible for high-resolution data. Compared to planar projections, the spherical approach has the advantage of avoiding distortions and preserving the isotropy of the signal, which is crucial for cosmological inference tasks \cite{gorski2005}.  In cosmological applications, DeepSphere has been applied to tasks such as weak lensing map classification, cosmological parameter estimation, and non-Gaussianity detection \cite{perraudin2019}. In our study, we extend its use to the problem of predicting five cosmological parameters --- the cold matter density $\omega_{\mathrm{c}}$, the baryon density $\omega_\mathrm{b}$, the amplitude $A_s$, the amplitude of the PMF $B_{1\rm{Mpc}}$, and the magnetic damping scale relation $\beta=\log_{10}(\tau_{\nu}/\tau_{\mathrm{B}})$, including PMF contributions directly from simulated CMB maps.

\subsection{Simulation of CMB maps with PMF}

We generated CMB temperature anisotropy maps using a modified pipeline based on \href{https://camb.info/}{CAMB} \cite{lewis2000,Shaw_2010}, the patch for CAMB that computes CMB sourced by PMFs~\cite{zucca2017}, and \href{https://healpix.sourceforge.io/}{Healpy} \cite{gorski2005}. This framework incorporates standard $\Lambda\mathrm{CDM}$ physics and assumes a flat universe. It is restricted to three massless neutrinos and passive, compensated magnetic modes that are mutually uncorrelated~\cite{zucca2017}. The total theoretical angular power spectrum is given by
\begin{equation}
\label{power_Spectra}
\mathrm{C}^{\text{theor}}_{l}=\mathrm{C}^{\text{prim}}_{l}+\mathrm{C}^{\text{pass}}_{{l}}+\mathrm{C}^{\text{comp}}_{l},
\end{equation}
which comprises three main components: the primordial (non-magnetic) spectrum from standard cosmological perturbations $\mathrm{C}^{\text{prim}}_{l}$, the passive magnetic contribution $\mathrm{C}^{\text{pass}}_{{l}}$ and the compensated magnetic contribution $\mathrm{C}^{\text{comp}}_{l}$. Scalar perturbations are presented in both, the compensated and passive modes, vector perturbation are included in the compensated contribution and tensor perturbation are included in the passive contribution. In this paper, we adopt the following notation: \textsc{full} to refer to the dataset including scalar, vector and  passive magnetic contributions, and \textsc{prim-vp} to refer to the dataset excluding the scalar compensated mode. Two independent datasets were constructed each with 1000 power spectra and 5 realizations, each dataset \textsc{full} and \textsc{prim-vp} contains 5000 maps. The cosmological parameters were set to Planck besfit cosmology~\cite{aghanim2020}, with the Hubble constant $H_0$ as $0.67556$, the scalar spectral index $n_{s}$ as $0.9665$ and the magnetic spectral index $n_{\mathrm{B}}$ as $-2.9$. Table \ref{param_ranges} summarizes the ranges of cosmological and primordial magnetic field (PMF) parameters employed in the generation of the datasets used in this study. These parameters span values chosen to cover the relevant physical scenarios considered in our analysis.

\begin{table}[H]
\centering
\begin{tabular}{lcc}
\hline
\textbf{Parameter} & \textbf{Min} & \textbf{Max} \\
\hline
$\omega_{\mathrm{c}}$        & 0.05     & 0.50 \\
$\omega_{\mathrm{b}}$      & 0.005        & 0.05 \\
$A_{s}$                        & $1.015 \times 10^{-9}$ & $4.015 \times 10^{-9}$ \\
$B_{\mathrm{1Mpc}}$       & 5        & 15 \\
$\beta=\log_{10}(\tau_{\nu}/\tau_{\mathrm{B}})$      & 5        & 20 \\
\hline
\end{tabular}
\caption{Ranges of cosmological and PMF parameters used for dataset generation. Here, $\omega_{\mathrm{c}}$ and $\omega_{\mathrm{b}}$ represent the cold dark matter and baryon density parameters, respectively, both dimensionless.
$A_s$ is the amplitude of the primordial scalar perturbations given in $\mathrm{Mpc}^{-3}$.
$B_{\mathrm{1Mpc}}$ denotes the magnetic field strength smoothed over 1 Mpc scale, expressed in nanogauss (nG).
$\beta=\log_{10}(\tau_{\nu}/\tau_{\mathrm{B}})$ is the logarithmic ratio of the magnetic damping scale, representing the ratio of the time $\tau_{\nu}$ of neutrino decoupling to the time $\tau_{\mathrm{B}}$ at which the magnetic field was generated in the early universe, a dimensionless parameter.}
\label{param_ranges}
\end{table}

\subsection{Dataset and Validation}

The CMB is modeled as a Gaussian random field, meaning that it is not a unique deterministic sky but a statistical ensemble of possible realizations, all characterized by the same angular power spectra $C_l$. While most values in these realizations cluster near the mean, a finite number of modes on the sky, especially at large angular scales where cosmic variance is high inevitably, leads to extreme fluctuations. These statistically plausible outliers are a natural consequence of cosmic variance  \cite{aghanim2020}.  The broad parameter ranges chosen for dataset generation (see Table~\ref{param_ranges})  were adopted following the analysis in~\cite{zucca2017} where they investigated constraints on PMF from Planck mission and South Pole Telescope data. Their choice of ranges is motivated by both observational and theoretical considerations. The damping scale ratio ($\beta=\log_{10}(\tau_{\nu}/\tau_{\mathrm{B}})$) for instance, can capture a variety of physically plausible PMF generation scenarios, including inflationary and phase-transition models. Standard cosmological parameters such as $\omega_{\text{c}}$, $\omega_{\text{b}}$ and $A_{s}$ were also varied within extended ranges compared to Planck priors. This approach is essential for exploring degeneracies between PMF effects and baseline cosmology. 
Furthermore, we implemented a data-cleaning procedure to remove maps with pixel values outside physically reasonable limits. The minimum and maximum pixel values of each image were examined, and a per-pixel threshold criterion was applied to flag anomalous maps. Maps exceeding these thresholds were classified as outliers and excluded from the dataset. Similar pre-processing and anomaly-rejection strategies have been employed in recent works applying deep learning to CMB data to ensure robust inference of cosmological parameters \cite{caldeira2019}.
We found extreme outliers in some maps with magnitudes up to $10^{30}$, which are effectively removed after thresholding, resulting in a clean and coherent data range below $10^{6}$ (more details in Appendix \ref{data_cleaning}). From the initial 5000 CMB maps generated with CAMB, \textsc{full} reaches 67 anomalous maps discarded, leaving 4933 valid maps, and \textsc{prim-vp} reaches 83 anomalous maps discarded, leaving 4917 valid maps. The cleaned dataset was randomly partitioned into training, validation, and test subsets using a two-step procedure. First, $70\%$ of the data was allocated to the training set, while the remaining $30\%$ was reserved for further splitting. This reserved portion was then evenly divided into validation and test sets, each comprising $15\%$ of the original dataset. In addition, we generated an independent dataset consisting of 400 simulated CMB maps with one realization, which serves as an external validation set for the model. This dataset was built under the same physical assumptions as the training data but with a distinct random seed. The main purpose of this external dataset is to evaluate the generalization capacity when exposed to unseen data distributions \cite{Kueppers2022}. With this, the test dataset measures the model's generalization during training, the validation dataset monitors overfitting and manages hyperparameter tuning, while external dataset assesses generalization to unseen data.

\subsection{Architecture}

The network architecture was inspired by the DeepSphere \emph{global regression} design \cite{perraudin2019} and implemented using \texttt{TensorFlow} with \texttt{HealpyGCNN} layers. It was designed to combine spherical convolutional feature extraction with Bayesian dense layers for probabilistic regression. The overall architecture consists of three main stages: normalization layer, hierarchical spherical convolution and pooling layers, and Bayesian dense layers for regression.

\begin{figure}[H]
    \centering
        \includegraphics[width=1\linewidth]{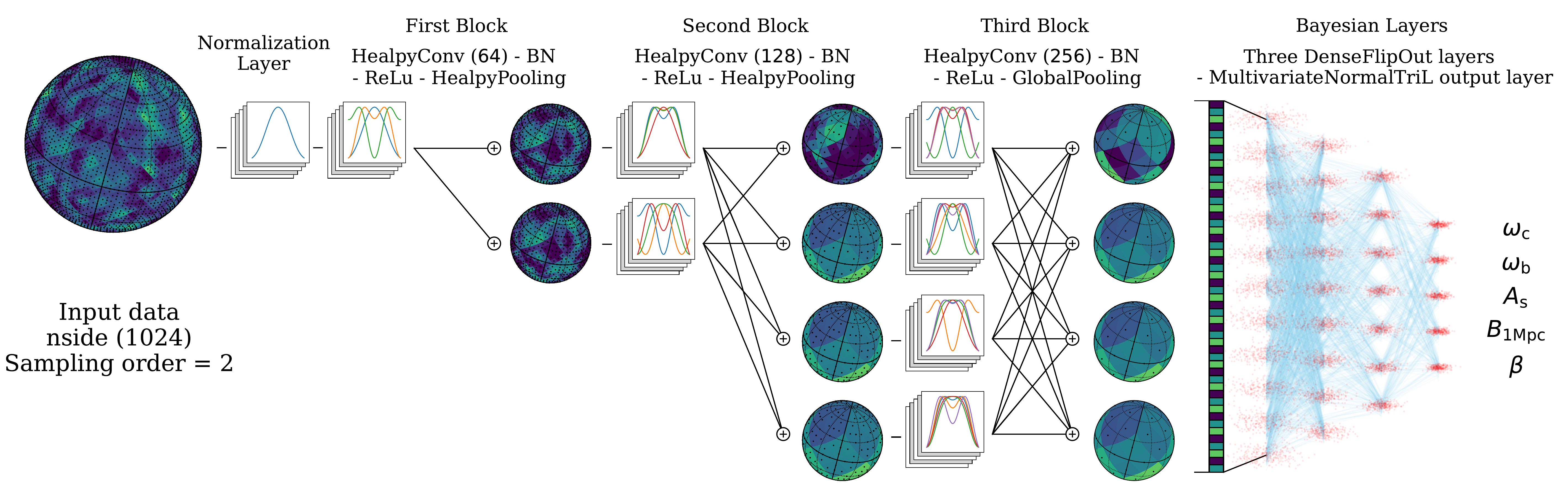}
    \caption{Representation of the neural network architecture used for cosmological parameter inference. 
    The input consists of full-sky spherical maps at resolution \texttt{nside = 1024}, which are first normalized using a normalization layer. 
    The architecture includes three convolutional blocks with \texttt{HealpyConv} layers, each followed by batch normalization (BN), ReLU activation, and average HealpyPooling layers. The third block has GlobalPooling. The extracted features are passed to Bayesian dense layers, implemented using \texttt{DenseFlipOut}, followed by a \texttt{MultivariateNormalTriL} output layer that predicts the posterior distribution of the target parameters: 
    cold dark matter density $\omega_{\mathrm{c}}$, baryon density $\omega_{\mathrm{b}}$, scalar amplitude $A_{s}$, magnetic field strength $B_{\mathrm{1Mpc}}$, and damping scale ratio $\beta = \log_{10}(\tau_{\nu}/\tau_{\mathrm{B}})$.}
    \label{fig: NN}
\end{figure}

The input
consists of full-sky spherical maps at resolution nside = 1024 and then, we create samples by dividing the complete spheres in patches (based on healpix sampling of order=2).  We applied an input normalization layer (N) based on the mean and variance of the training dataset by iterating over all input maps. These statistics were stored and later used to initialize the \href{https://keras.io/api/layers/preprocessing_layers/numerical/normalization/
}{\texttt{Normalization} layer} in Keras that standardizes the inputs. We employed a sequence of \texttt{HealpyPseudoConv} and \texttt{HealpyPool} layers. These layers exploit the hierarchical structure of the \texttt{HEALPix} pixelization. Each block contains a \texttt{HealpyPseudoConv} layer (GC)  with $p=1$ and an increasing number of output channels ($F_\mathrm{out}=64, 128, 256$), a batch normalization layer (Bn), and a ReLU activation.  Between blocks, we introduced spherical pooling layers (\texttt{HealpyPooling}, P) with average reduction of the resolution (\texttt{nside}). The outputs of the last spherical convolutional block are aggregated using a global average pooling layer. This reduces the feature maps to a compact vector representation that can be passed to fully connected layers. To incorporate predictive uncertainty, we employed three \texttt{DenseFlipout} layers \cite{wen2018flipout} from TensorFlow Probability with $256$, $128$, and $64$ units, each using a ReLU activation. The Bayesian treatment is achieved through variational inference: weights are modeled as distributions with a posterior approximated by the Flipout estimator. The final regression layer is a Bayesian multivariate distribution. Specifically, we used \href{https://www.tensorflow.org/probability/api_docs/python/tfp/layers/DenseFlipout}{\texttt{DenseFlipout}} to output the parameters of a \href{https://www.tensorflow.org/probability/api_docs/python/tfp/distributions/MultivariateNormalTriL}{\texttt{MultivariateNormalTriL} }distribution \cite{tran2018bayesianlayers} of dimension five (corresponding to $\omega_{\mathrm{c}}$, $\omega_{\mathrm{b}}$,  $A_s$, $B_{1Mpc}$, and $\beta=\log_{10}(\tau_{\nu}/\tau_{\mathrm{B}})$).

\subsection{Training}

The regression network was trained to predict the five cosmological parameters from simulated CMB maps using a probabilistic approach. The input shape was defined as $(32, N_{\mathrm{pix}}, 3)$, where $32$ is the number of samples, $N_{\mathrm{pix}}$ denotes the number of padded pixels per map and the factor of three corresponds to the channels of the simulation. 
The model was compiled with the Adam optimizer, using a learning rate of $1 \times 10^{-4}$, and the negative log-likelihood ($\texttt{NLL}$) along with the (KL) was employed as the loss function. 
In addition, mean squared error (\texttt{mse}) and mean absolute error (\texttt{MAE}) were monitored as complementary metrics.\\

Regarding the callbacks, three training callbacks were included: (1) \texttt{EarlyStopping} with a patience of 25 and automatic restoration of the best weights, (2) \texttt{ModelCheckpoint} for saving the optimal weights during training, and (3) \texttt{ReduceLROnPlateau}, which decreased the learning rate by a factor of 0.6 after five consecutive non-improving validation steps, with a floor of $10^{-6}$. These strategies have been widely adopted in deep learning to prevent overfitting and to stabilize optimization in non-convex loss landscapes \cite{goodfellow2016}. Training was performed for 300 epochs with a batch size of 32. The total number of trainable parameters amounted to approximately $3.8 \times 10^{5}$. The architecture combined Healpy-based convolutional layers for feature extraction with Bayesian dense layers for uncertainty-aware regression, resulting in an end-to-end model capable of mapping spherical CMB realizations to cosmological parameters.

\subsection{Uncertainty Calibration}
Uncertainty in deep neural networks tend to be (under)over-confident in their predictions. The ways to diagnose the  uncertainty quality is through reliability plots~\cite{guo2017}. Calibrating neural networks during training is not efficient in most cases because it affects the model performance. However, after training methods, known as post hoc methods, preserve the precision achieved during the training period~\cite{hortua2020}. Here we leverage  post-hoc calibration to improve the reliability of the uncertainty estimates. In the context of Bayesian regression, well calibrated uncertainty estimates are essential for ensuring that predictive intervals reflect true empirical coverage. We employed the~\href{https://github.com/EFS-OpenSource/calibration-framework}{NetCal} calibration framework, which has been extended to support probabilistic regression tasks. These techniques aim to adjust the predictive distributions outcomes such that the nominal coverage of predictive intervals aligns with their observed coverage. In fact, we implemented  two post-hoc calibrations methods introduced in~\cite{Kueppers2022}, the Variance Scaling and GPNormal.  The \texttt{VarianceScaling} method applies a correction factor, $\alpha$, to rescale the predictive variance $\sigma^{2}_{\text{cal}} = \alpha\cdot\sigma^{2}_{\text{pred}}$ and re-calibrate the uncertainties. GPNormal, on the other hand, uses a Gaussian process to map the predictive mean and variance to calibrated uncertainty estimates, allowing for more flexible non-linear transformation of the uncertainty space~\cite{Kueppers2022}.
\section{Results \label{sec:4}}

In this section, we report the results obtained using two simulated datasets that differ in their scalar components: \textsc{full} and \textsc{prim-vp}. Both datasets were used to train and evaluate the same neural network model described previously, with the goal of predicting the five cosmological parameters listed in Table \ref{param_ranges}. To assess the model's performance, we evaluated it on the respective test sets and computed two key metrics:  mean squared error (\texttt{MSE}), and mean absolute error (\texttt{MAE}) along with the  loss function described by Eq. \ref{nll}. Table \ref{loss_comparison} summarizes the evaluation results for both datasets:

\begin{table}[H]
\centering
\begin{tabular}{lcccccc}
\toprule
\multirow{2}{*}{Set} & 
\multicolumn{3}{c}{\textbf{DB-nmsvp}} & 
\multicolumn{3}{c}{\textbf{DB-nmvp}} \\
\cmidrule(lr){2-4} \cmidrule(lr){5-7}
 & Loss & MSE & MAE & Loss & MSE & MAE \\
\midrule
Validation & 5250.468 & 0.055 & 0.154 & 5254.806 & 0.066 & 0.174 \\
Test       & 5250.208 & 0.050 & 0.147 & 5254.857 & 0.068 & 0.176 \\
External   & 5250.396 & 0.052 & 0.151 & 5255.045 & 0.068 & 0.176 \\
\bottomrule
\end{tabular}
\caption{Comparison of performance (\texttt{Loss}, \texttt{MSE}, and \texttt{MAE}) evaluated on the validation, test and external sets of both datasets. A lower value indicates better performance.}
\label{loss_comparison}
\end{table}

We observe that the model reaches comparable training stability across both datasets. However, the \textsc{full} outperformed the \textsc{prim-vp} dataset across all three metrics.  Since \texttt{MSE} penalizes larger errors more heavily, its lower value shows that the model was more effective at suppressing outlier predictions when trained with the full scalar input. These results demonstrate that the scalar input in the \textsc{full} dataset enhances the model's capacity to learn the underlying mapping between CMB features and cosmological parameters, including the effects of primordial magnetic fields. 

\begin{table}[H]
\centering
\begin{tabular}{lcc}
\toprule
\textbf{Parameter} & \textbf{\textsc{FULL}} & \textbf{\textsc{PRIM-VP}} \\
\midrule
$\omega_{c}$            & 0.773 & 0.671 \\
$\omega_{b}$       & 0.985 & 0.973 \\

$A_{s}$            & 0.626 & 0.562 \\
$B_{1\mathrm{Mpc}}$       & 0.992 & 0.982 \\
$\beta=\log_{10}(\tau_{\nu}/\tau_{\mathrm{B}})$ & 0.895 & 0.856 \\
\bottomrule
\end{tabular}
\caption{$R^2$ scores on the test set for each predicted cosmological parameter. The models correspond to \textsc{full} (and passive components) and \textsc{prim-vp}.}
\label{r2_test}
\end{table}
Table \ref{r2_test} reports the $R^2$ for each parameter in both experiments. Baryon density and the magnetic field are the best-predicted parameters in both datasets with high precision, while the magnetic damping scale obtained the lowest value, especially in the \textsc{prim-vp} case, which lacks the scalar component. Some features useful for predicting $\beta$ are encoded in the scalar contribution of the CMB maps.

\begin{figure}[H]
    \centering
    \includegraphics[width=1\linewidth]{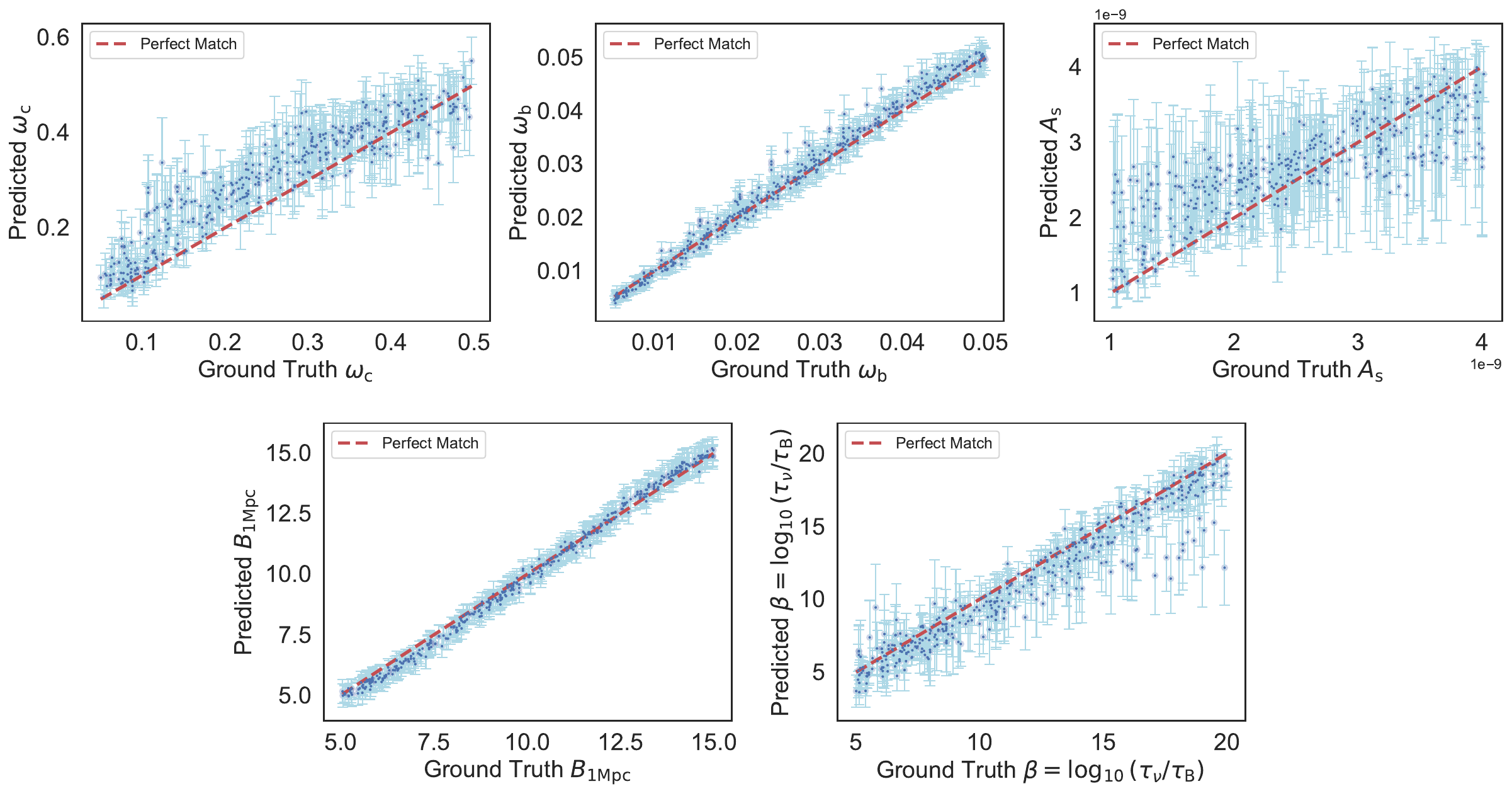}
    \caption{Predicted versus ground truth true values for the five cosmological parameters using the \textsc{full}. Each subplot corresponds to one parameter, with the black dashed line representing the ideal prediction. Shaded bars denote one standard deviation.}
    \label{r2_DB-nmsvp}
\end{figure}

\begin{figure}[H]
    \centering
    \vspace{0.5cm}
    \begin{subfigure}[b]{0.32\textwidth}
        \includegraphics[width=\textwidth]{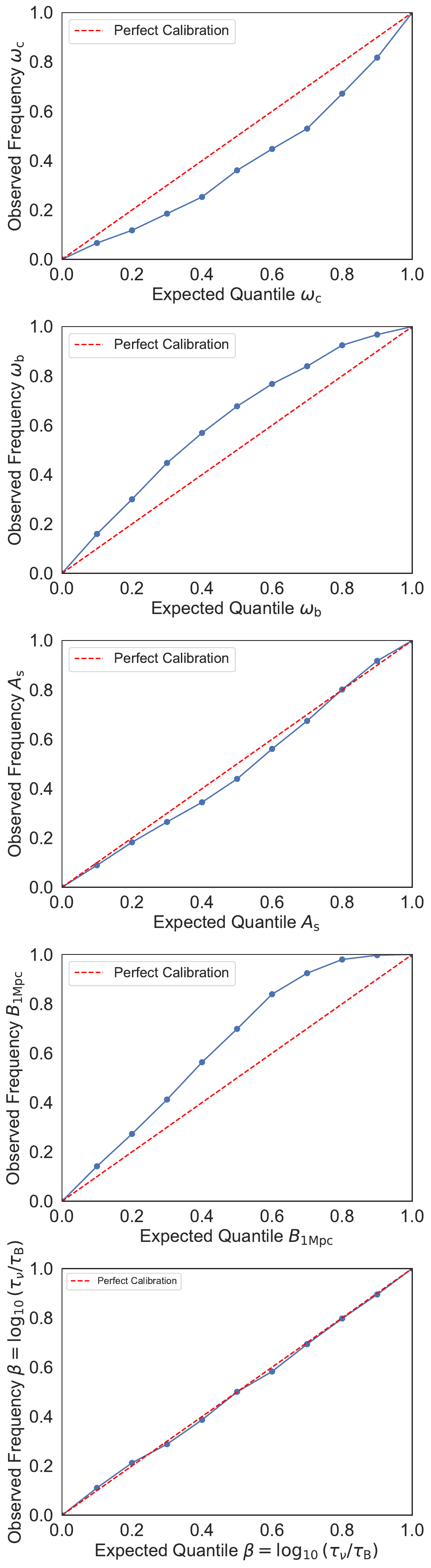}
        \caption{Uncalibrated}
    \end{subfigure}
    \hfill
    \begin{subfigure}[b]{0.32\textwidth}
        \includegraphics[width=\textwidth]{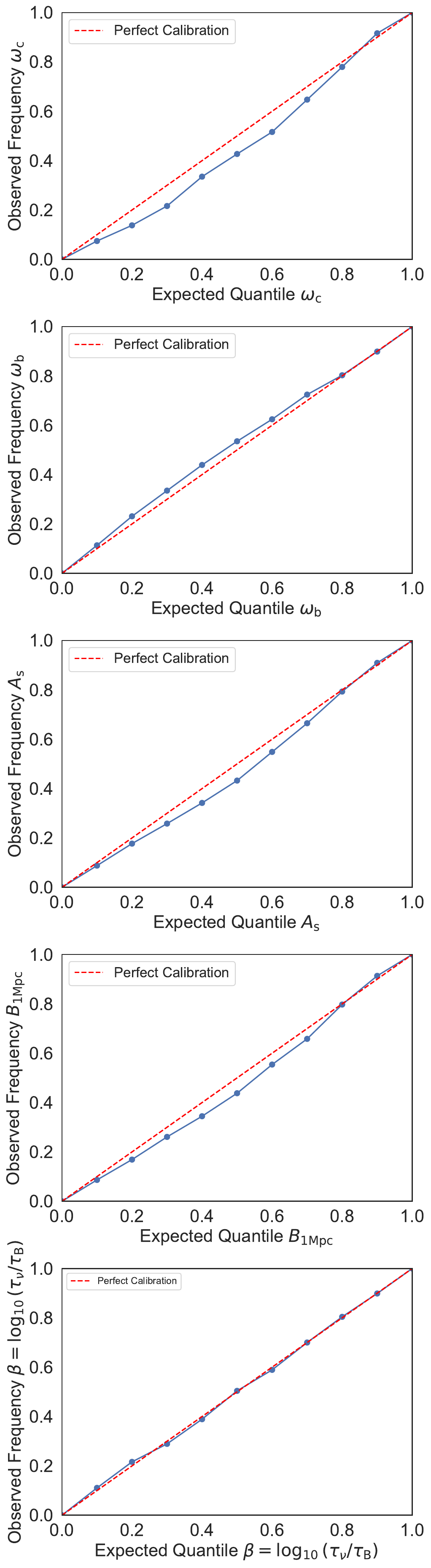}
        \caption{Variance Scaling}
    \end{subfigure}
    \hfill
    \begin{subfigure}[b]{0.32\textwidth}
        \includegraphics[width=\textwidth]{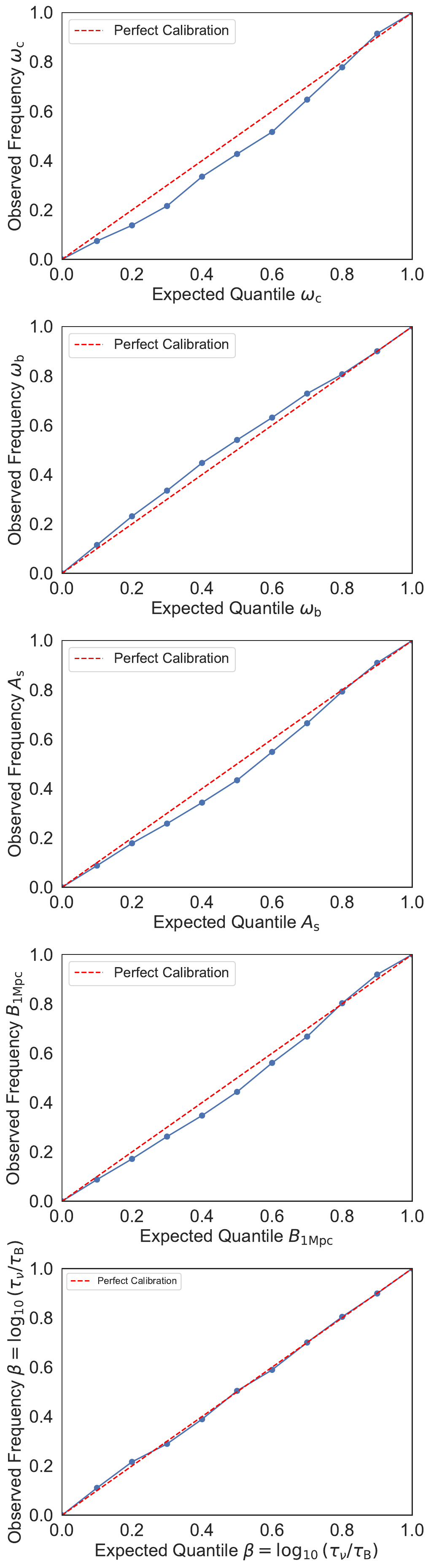}
        \caption{GP Normal}
    \end{subfigure}
    \caption{Uncertainty calibration curves (reliability diagrams) for each predicted parameter using the \textsc{full} dataset. The plots compare predicted Excepted quantile (x-axis) with the observed frequencies (y-axis). Perfect calibration corresponds to the dashed red diagonal. \textbf{(a)} Uncalibrated model outputs, \textbf{(b)} Post-hoc calibration using \texttt{VarianceScaling}. \textbf{(c)} Post-hoc calibration using \texttt{GPNormal}.}
    \label{reliability_curves}
\end{figure}

Figure~\ref{r2_DB-nmsvp} shows the predicted values versus the ground truth for each of the five parameters in the \textsc{full} dataset. The red dashed line represents the perfect match. Most predictions align well along this line, especially for $B_{1\mathrm{Mpc}}$ and $\omega_b$, confirming the strong performance seen in the $R^2$ scores. The error bars represent the predicted standard deviation of the output distribution.  Parameters with lower $R^2$ tend to have larger uncertainty, indicating the model is also less confident when predictions are less accurate. Nevertheless, the resulting uncertainty estimates may still be uncalibrated due to the model misspecification or approximation errors \cite{Kuleshov2018}. In Fig.~\ref{reliability_curves} we display the reliability diagrams when we use VarianceScaling and GPNormal calibrations for the five output dimensions. Each plot compares the expected quantile against the observed frequency.

Regarding the uncalibrated curves (see Fig.~\ref{reliability_curves}.a), the parameters with the highest $R^2$ scores — $\omega_b$ and $B_{1\mathrm{Mpc}}$ — show clear overconfidence from 0.2 to 0.8 in the expected quantile. In $B_{1\mathrm{Mpc}}$ the overconfidence is more remarkable in quantile 0.6. The predicted confidence intervals are too narrow and capture fewer true values than expected, even when point predictions are highly accurate, the associated uncertainty estimates can still be miscalibrated. On the other hand, the parameters $A_s$ and $\beta = \log_{10}(\tau_\nu / \tau_B)$, which have lower $R^2$ scores, show well-calibrated, even slightly underconfident behavior of $A_s$ in quantile 0.5, resulting in curves that lie above the diagonal. After applying post-hoc calibration (see Fig.~\ref{reliability_curves}.b-c), both \texttt{VarianceScaling} and \texttt{GPNormal} improve the alignment between predicted confidence levels and empirical frequencies, thereby reporting reliable uncertainties for our results. After applying the calibration procedure, it is important to fine-tune the model incorporating the recalibration adjustments to ensure that these corrections are consistent with the learned representations. Otherwise, post-hoc corrections may lead to a mismatch between the adjusted uncertainty estimates and the actual error distribution. This principle has been discussed in the recent literature, where integrating calibration directly into the training loop has been shown to improve both reliability and generalization \cite{dheur2024}.
\section{Discussion \label{sec:5}}
We discuss the main experimental and architectural challenges found during the implementation of the Bayesian graph neural network. The computational process was executed using two NVIDIA TESLA V100-SXM2 GPUs running on the Google Cloud Platform.
The initial architecture suggested by the main \texttt{DeepSphere paper}\cite{perraudin2019} --i.e, $NN = AV\circ GC\circ\sigma\circ GC\circ P\circ\sigma\circ BN\circ GC$ proved effective for predicting a single cosmological parameter. The normalization layer showed instability and tended toward infinite values because of large outputs in the input data. After the data cleaning phase to remove these outliers, the normalization layer stabilized and assumed reliable values. The project initial promising results with the prediction of a single cosmological parameter led to the generation of a new dataset designed to predict two parameters by mixing different components of the \texttt{PMF}, but failed when scaled the architecture to simultaneously predict two or more parameters. To address this issue, the architecture was modified by adding a similar block to the original structure, and the number of convolutional filters was increased. Regarding the pooling operation, experimenting with both Average and Maximum strategies revealed a performance dependency on the output size. The Max pooling yielded better results for predicting a single parameter, whereas AVG pooling provided superior performance for the simultaneous prediction of multiple cosmological parameters. Subsequent success with two parameters lead to the creation of datasets for three parameters and then four parameters. In this scenario, three dense layers were applied instead of two, for better performance.  We identified and selected the two best performing datasets, \textsc{full} and \textsc{prim-vp}, chosen based on preliminary results that showed the highest $R^{2}$ scores across the four parameters (see table \ref{deter_model_r2} of Appendix A) which were then used to generate the final datasets for the prediction of the cosmological parameters. The training was conducted using a batch size of 32 and a learning rate of $1\times10^{-4}$. The training time was on average 8 second per step for 108 steps per epoch, resulting in approximately 860 seconds per epoch. In contrast to the deterministic model, the probabilistic model exhibited superior training stability and did not require stabilization techniques such as \texttt{ReduceLROnPlateau} or \texttt{EarlyStopping} to avoid divergence. The final model has $382868$ parameters. We have published a subset of the dataset on \href{https://doi.org/10.5281/zenodo.17400503}{\texttt{Zenodo}} for public access and further use, and the scripts used in this analysis are publicly available in the following repository~\href{https://github.com/JavierOrjuela/Bayesian-SphericalCNN}{\faGithub}.
\section{Conclusions}\label{section6}
In this work, we have introduced a novel Bayesian graph deep learning framework for estimating key cosmological parameters from simulated full-sky CMB maps, explicitly incorporating contributions from PMFs through their passive and vector modes. By extending the \texttt{DeepSphere} spherical convolutional neural network with Bayesian neural network layers, our approach simultaneously leverages the geometry of spherical data and provides rigorous uncertainty quantification. The main conclusions of our study can be summarized as follows:
\begin{enumerate}
    \item[(i)] High predictive accuracy through integration of physical characteristics.
The proposed \texttt{B-GNN} model achieved $R^2$ scores close to $0.9$ for most of the five cosmological parameters considered: $\omega_b$, $\omega_c$, $A_s$, $B_{1\mathrm{Mpc}}$, and $\beta = \log_{10}(\tau_\nu/\tau_B)$. We found that the inclusion of PMF-induced passive and vector modes was crucial for improving the model’s representational power. This is clearly reflected in the superior performance of the \textsc{full} dataset compared to \textsc{prim-vp}, demonstrating that embedding relevant physical information in the input enhances the accuracy and stability of the predictions.
    \item[(ii)] Reliable and calibrated uncertainty quantification.
A central contribution of this work is the incorporation of predictive uncertainty estimates. Although the raw uncertainty outputs of the probabilistic model exhibited overconfidence, applying post-hoc calibration methods such as \texttt{VarianceScaling} and \texttt{GPNormal} effectively corrected this behavior. The resulting calibrated confidence intervals accurately matched the empirical error distribution, thereby increasing the interpretability and reliability of the model’s predictions. This is particularly important in cosmology, where well-calibrated uncertainties are essential for meaningful parameter inference.
    \item[(iii)] Strong generalization and architectural robustness.
The model showed stable performance across independent validation datasets, confirming its ability to generalize beyond the training distribution. This robustness is the result of a carefully designed training strategy, including regularization, optimal learning rate scheduling, and dropout. Importantly, unlike the deterministic baseline, the Bayesian model maintained stability without requiring additional early stopping or learning-rate reduction techniques, further supporting its suitability for real data applications.
    \item [(iv)] Advancement of cosmological parameter inference with Bayesian deep learning.
Compared to traditional parameter estimation techniques, this framework integrates a spherical CNN architecture with Bayesian inference principles, enabling both accurate point predictions and principled uncertainty estimation within a single model.
    \item[(v)] Outlook and implications for future applications.
Our results indicate that a well-calibrated Bayesian neural network can robustly infer cosmological parameters from complex, high-dimensional CMB data that include subtle physical effects such as primordial magnetism. This framework offers a flexible and powerful alternative to conventional inference pipelines, providing not only point estimates but also interpretable uncertainty information. These characteristics make it a promising tool for future analyses of real full-sky CMB observations, especially in the context of upcoming high-resolution surveys.
\end{enumerate}

\noindent In summary, this study demonstrates that probabilistic deep learning methods, when carefully designed and systematically calibrated, can significantly enhance the precision and reliability of cosmological parameter estimation from CMB data. The integration of physical information, spherical geometry, and Bayesian inference principles offers a new and powerful pathway to tackle open questions in cosmology and to explore beyond-standard physics.

\begin{acknowledgements}
This paper is based on work supported by the Google Cloud Research Credits program with the award GCP19980904
\end{acknowledgements}
\bibliographystyle{unsrtnat}
\bibliography{sections/Bibliography.bib} 

\section*{Appendix \label{appex}}
\subsection{Selection of best results \label{best_results}}
We chose these two dataset \textsc{full} and \textsc{prim-vp} due to preliminary results with a deterministic model and 4 cosmological predictable parameters:

\begin{table}[H]
\centering
\begin{tabular}{lccccc}
\toprule
\textbf{Datasets} & $\langle R^2\rangle$ & $\omega_c$ & $A_s$ & $B_{1Mpc}$ & $\beta = \log_{10}(\tau_\nu / \tau_B)$ \\
\midrule
\textsc{full} & \textbf{0.894} & 0.911 & 0.758 & 0.993 & 0.913 \\
\textsc{prim-vp}  & \textbf{0.880} & 0.891 & 0.842 & 0.923 & 0.863 \\
\textsc{svp}   & 0.860 & 0.876 & 0.660 & 0.992 & 0.912 \\
\textsc{vp}    & 0.800 & 0.959 & 0.258 & 0.996 & 0.985 \\
\textsc{prim-sv}  & 0.783 & 0.920 & 0.956 & 0.977 & 0.281 \\
\textsc{prim-v}   & 0.775 & 0.951 & 0.980 & 0.990 & 0.179 \\
\textsc{sv}    & 0.541 & 0.955 & 0.069 & 0.992 & 0.148 \\
\textsc{prim-s}   & 0.229 & 0.351 & 0.888 & -0.119 & -0.206 \\
\textsc{sp}    & 0.209 & -0.040 & -0.102 & 0.725 & 0.251 \\
\textsc{prim-sp}  & 0.054 & -0.011 & 0.043 & 0.181 & 0.005 \\
\bottomrule
\end{tabular}
\caption{Coefficient of determination ($R^2$) for the prediction of four cosmological parameters across all simulated datasets. 
Each column corresponds to one target parameter: the cold dark matter density ($\omega_c$), the scalar amplitude ($A_s$), 
the magnetic field strength smoothed at $1\,\mathrm{Mpc}$ ($B_{1\mathrm{Mpc}}$), and magnetic damping scale 
$\beta = \log_{10}(\tau_\nu / \tau_B)$. The first column reports the mean $\langle R^2 \rangle$ over all parameters.}
\label{deter_model_r2}
\end{table}

These results helped us to determine the two best datasets for modifying the dense layers with Bayesian layers, thereby obtaining more reliable results thanks to the implementation of uncertainty quantification. The deterministic model performed better results of $R^{2}$ in \textsc{full} and \textsc{prim-vp}, achieving $0.894$ and $0.88$, respectively. In contrast, the remaining datasets show notably lower $R^2$ scores, which suggests that the absence of certain perturbation modes limits the model’s ability to recover the full parameter space. This pattern highlights the importance of including multiple mode contributions in the simulated CMB maps. In particular, datasets that include the vectorial and passive (\textsc{vp}) modes consistently yield the highest correlations, indicating that these components encode essential physical information that improves the model’s capacity to recover the cosmological parameters. After selecting the two best datasets and configuring the probabilistic model as \texttt{B-GNN}, we generated the corresponding datasets with five cosmological parameters to be predicted and applied a data-cleaning procedure.
\subsection{Data Cleaning \label{data_cleaning}}
The data cleaning process corresponds to drop the outliers from the dataset. These outliers have huge values compared with the central tendency of the \texttt{CMB} maps. We present the comparison of the dataset before and after the cleaning process.

\begin{figure}[H]
    \centering
    \begin{subfigure}[b]{1\textwidth}
\includegraphics[width=1\linewidth]{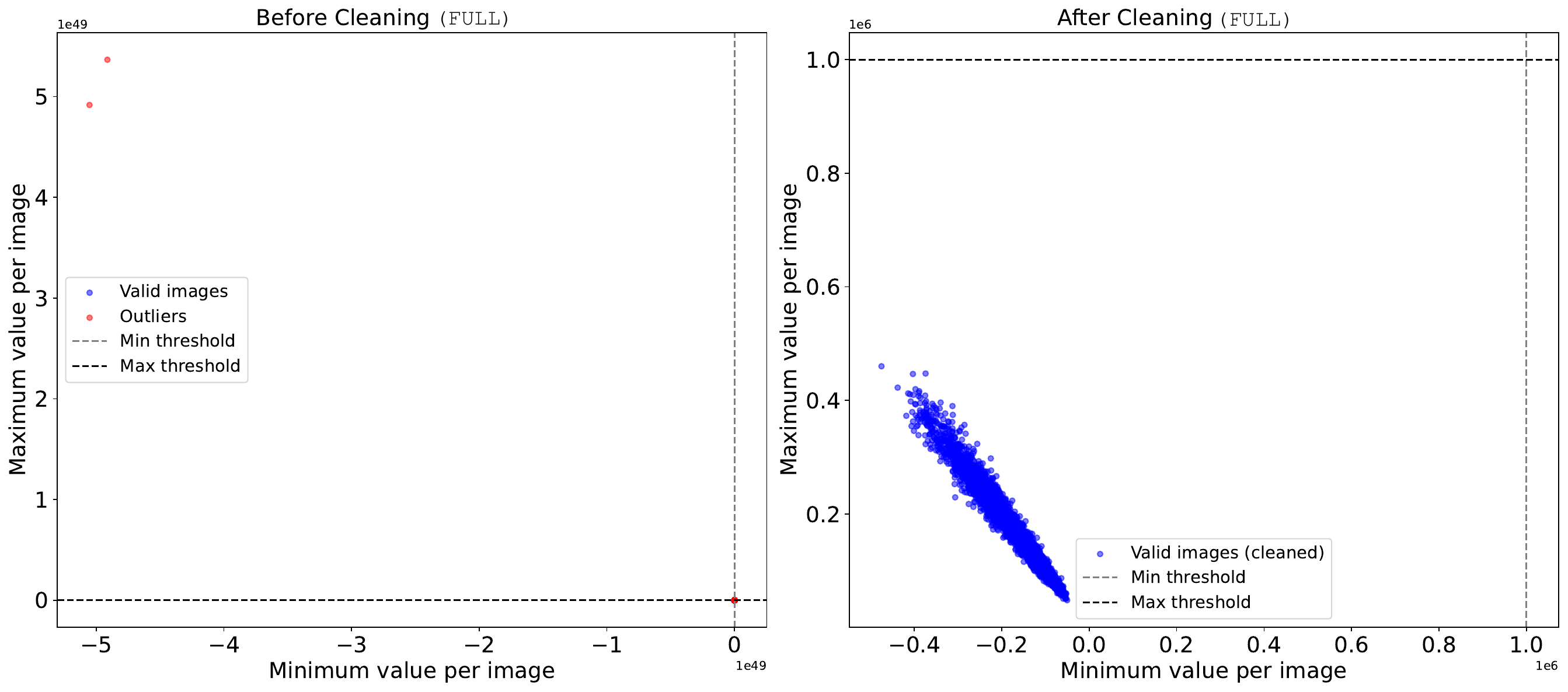}
        \caption{\textsc{full}}
    \end{subfigure}
    \hfill
    \begin{subfigure}[b]{1\textwidth}
\includegraphics[width=1\linewidth]{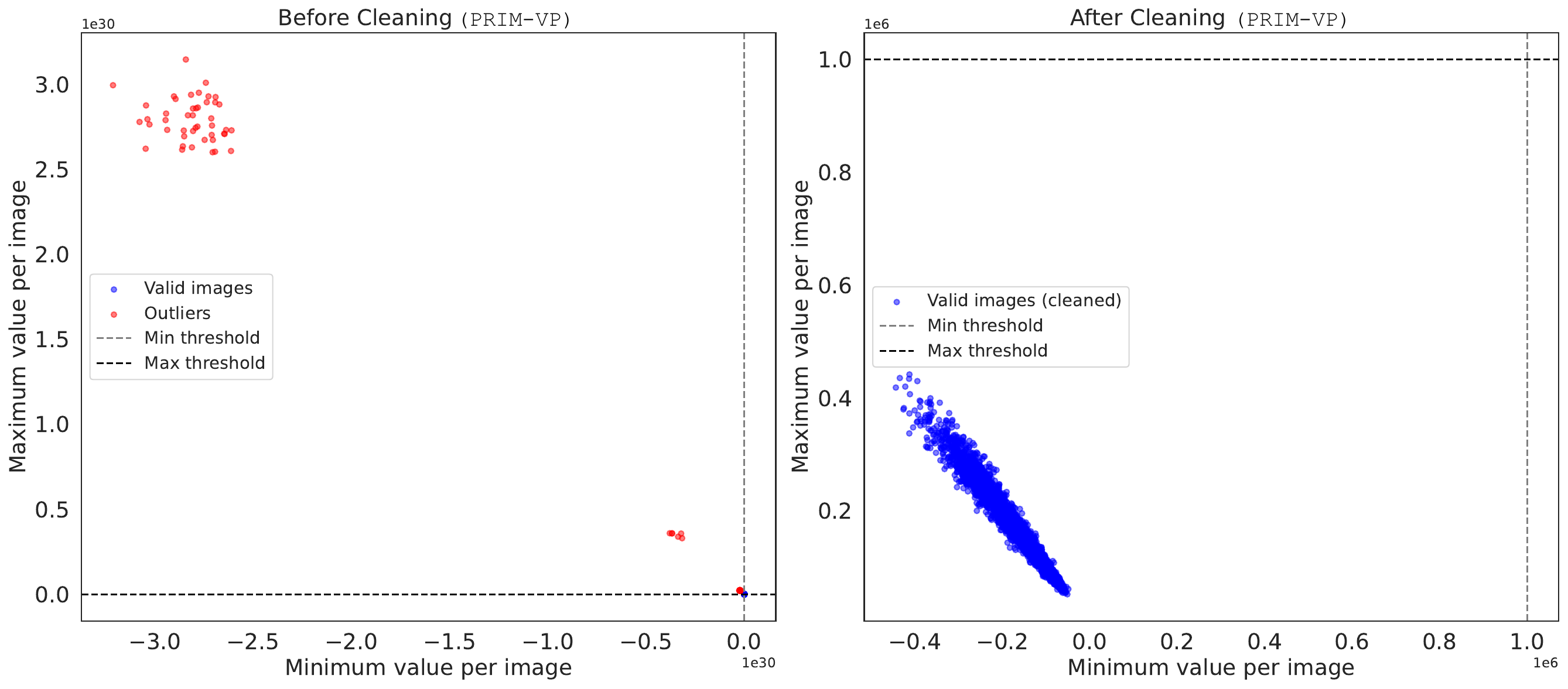}
        \caption{\textsc{prim-vp}}
    \end{subfigure}
    \hfill
    \caption{Scatter plots of the minimum vs. maximum pixel values per image before and after outlier removal. The left panel shows the raw data, where extreme outliers reach values up to $1\times10^{30}$. Panels (a) correspond to the \textsc{full} dataset, and panels (b) to the \textsc{prim-vp} dataset. The left panels show the distribution of the minimum and maximum pixel values per image before cleaning. The right panel displays the cleaned dataset after applying the thresholds with value $1\times10^{6}$. Dashed lines indicate the minimum and maximum thresholds used for cleaning.}
    \label{cleaning}
\end{figure}

The scatter plots in Figure \ref{cleaning} correspond to the data cleaning process in \textsc{full} and \textsc{prim-vp} dataset, illustrating the distribution of the minimum and maximum pixel values per image before and after applying the outlier cleaning procedure. The data cleaning procedure removes extreme outliers present in the simulation of the \texttt{CMB} maps. These anomalies were clearly visible because of the huge difference in scale comparing the left panel (before cleaning) and the right panel (after cleaning). The left panels present a scale of $1\times10^{49}$ while the right panels present a scale of $1\times10^{6}$, The presence of these outliers far far away of the main points affected the normalization step, driving the feature scaling to divergent values and destabilizing the training process. After removal, both datasets showed consistent range, as observed in the right panels. The cleaning step were crucial to guarantee numerical stability and reliable model calibration.

\subsection{Supplementary \textsc{prim-vp} results \label{suppl_results}}

While the main purpose of this paper focused in \textsc{full} results, an analysis of \textsc{prim-vp} results is presented. We report the $R^{2}$ of the five cosmological parameters. 

\begin{figure}[H]
    \centering
    \includegraphics[width=1\linewidth]{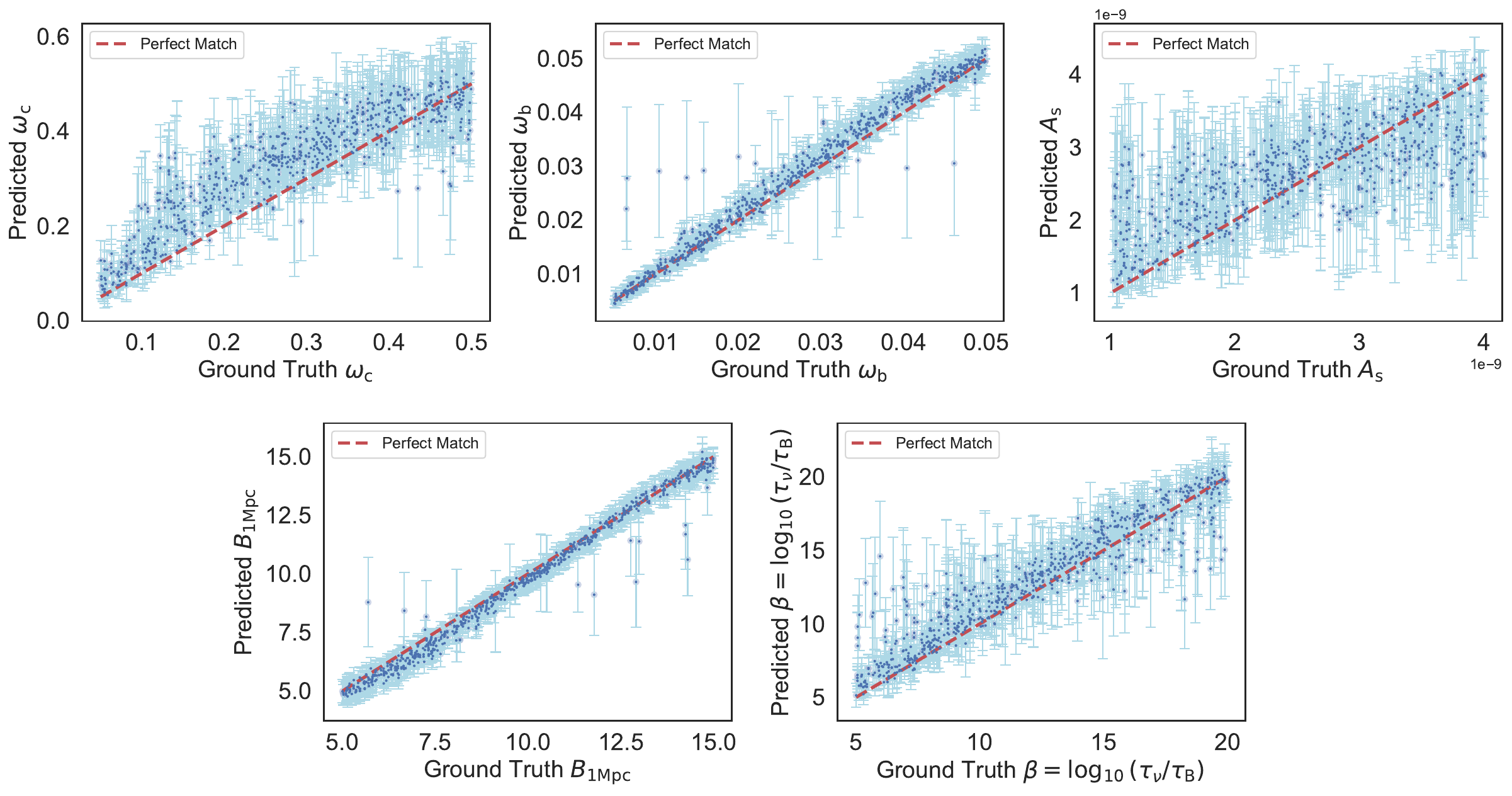}
    \caption{Predicted versus ground truth true values for the five cosmological parameters using the \textsc{prim-vp}. Each subplot corresponds to one parameter, with the black dashed line representing the ideal prediction. Shaded bars denote one standard deviation.}
    \label{r2_DB-nmvp}
\end{figure}

Among the parameters, $\omega_b$ and $B_{1,\mathrm{Mpc}}$ display the most accurate constrained predictions. In contrast, $A_s$ and $\beta$ show slightly higher scatter and wider uncertainty ranges, these parameters are more degenerated in the input features. Nonetheless, the overall alignment between predicted and true values confirms that the probabilistic model effectively captures both the mean trends and the associated uncertainties of the cosmological parameters.The large standard deviation for certain points out of the diagonal, particularly pronounced in $\Omega_{b}$ and $\beta=\log_{10}(\tau_{\nu}/\tau_{\mathrm{B}})$, is most likely due to these maps lying in marginal regions of the parameter manifold, which results in a wide predictive posterior distribution fron the \texttt{B-GNN} due to less training samples.

\begin{figure}[H]
    \centering
    \vspace{0.5cm}
    \begin{subfigure}[b]{0.32\textwidth}
        \includegraphics[width=\textwidth]{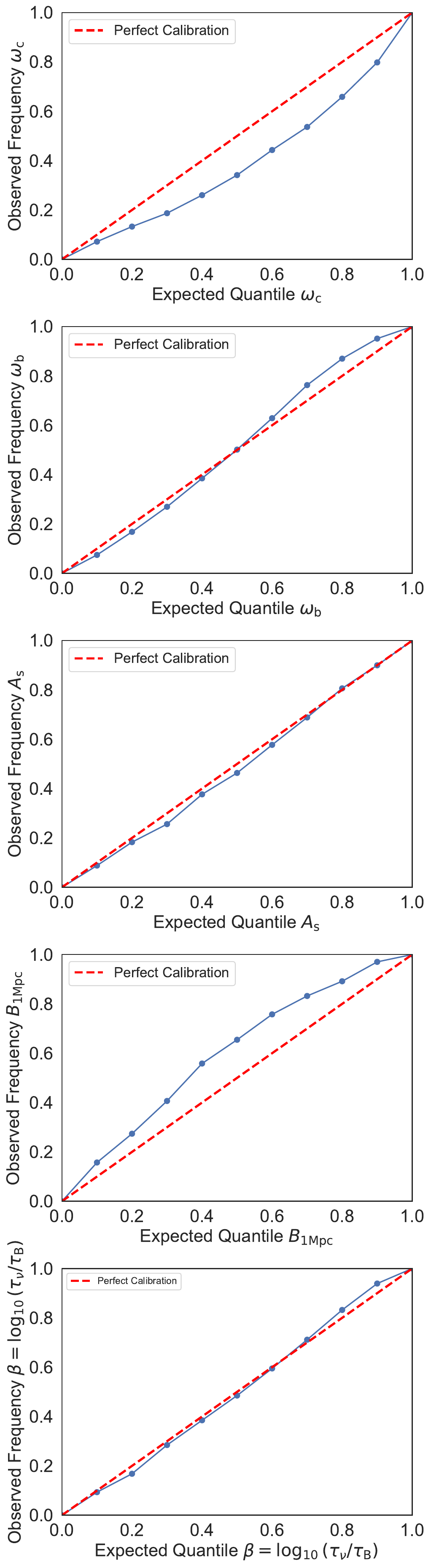}
        \caption{Uncalibrated}
    \end{subfigure}
    \hfill
    \begin{subfigure}[b]{0.32\textwidth}
        \includegraphics[width=\textwidth]{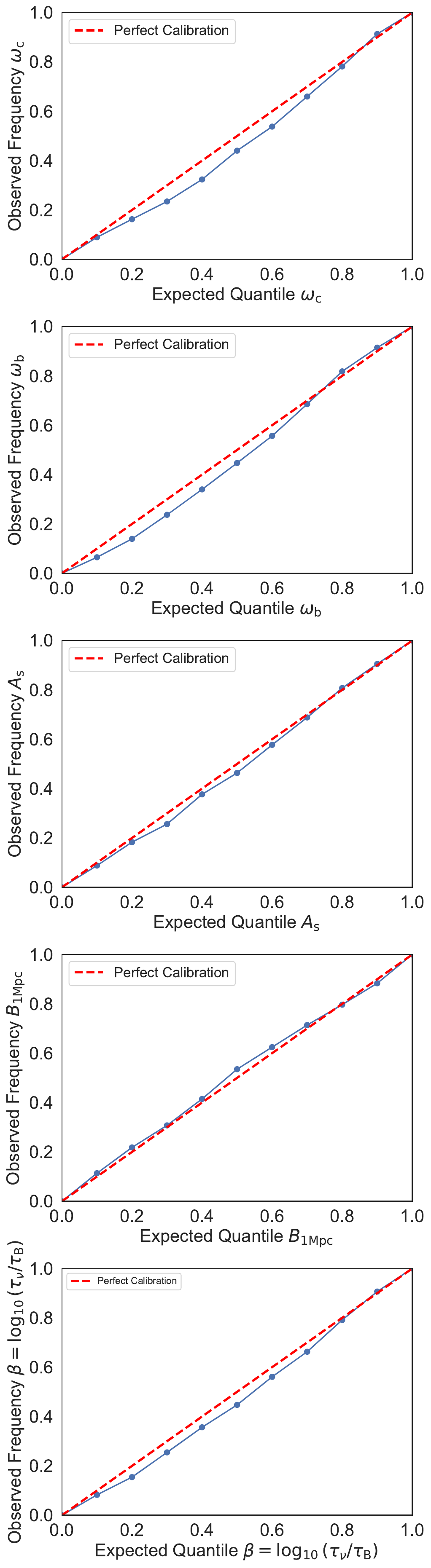}
        \caption{Variance Scaling}
    \end{subfigure}
    \hfill
    \begin{subfigure}[b]{0.32\textwidth}
        \includegraphics[width=\textwidth]{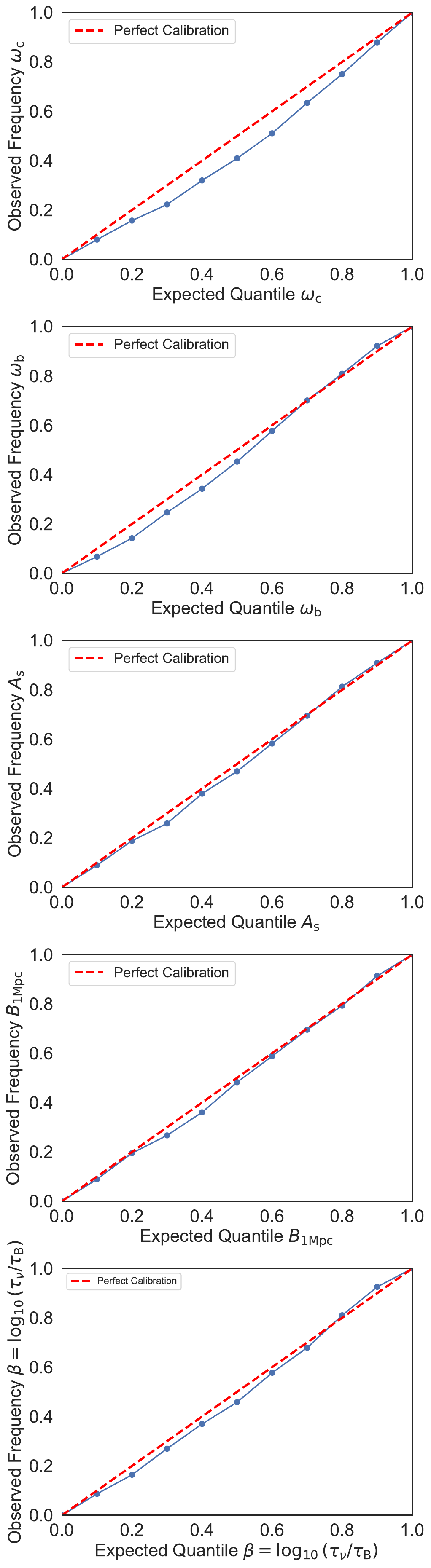}
        \caption{GP Normal}
    \end{subfigure}
    \caption{Uncertainty calibration curves (reliability diagrams) for each predicted parameter using the \textsc{prim-vp} dataset. The plots compare predicted Excepted quantile (x-axis) with the observed frequencies (y-axis). Perfect calibration corresponds to the dashed red diagonal. \textbf{(a)} Uncalibrated model outputs, \textbf{(b)} Post-hoc calibration using \texttt{VarianceScaling}. \textbf{(c)} Post-hoc calibration using \texttt{GPNormal}.}
    \label{cal_appex}
\end{figure}

Also, we performed the quality of the parameter uncertainties and overall calibration of the results. The figure \ref{cal_appex} displays reliability diagrams comparing the observed versus expected frequencies of the predicted uncertainty quantiles, allowing assessment of the probabilistic calibration of the model. The red dashed line represents perfect calibration, where predicted confidence intervals exactly match the empirical coverage.  The model for $B_{1Mpc}$ and $\Omega_{c}$ parameters, in figure \ref{cal_appex} (a), are overconfidence and underconfidence, respectively. In the other parameters there is a favorable agreement between the predicted  and the observed confidence levels. To reduce the miscalibration, we applied VarianceScaling(figure\ref{cal_appex}(b)) and GPNormal (figure\ref{cal_appex}(c)), reflecting an improved alignment between predicted and observed probabilities. This behavior demonstrates that the post-hoc calibration procedures corrected the mismatch between nominal and empirical coverage, leading to well-calibrated predictive distributions.

\end{document}